\documentclass{epl}

\def\eps{\epsilon} 
\def\ltsim{\hbox{ 
  \vbox to13pt{ 
  \vfill 
  \kern5.5pt 
  \hbox{$<$} 
  \kern-15pt 
  \hbox{$\sim$} 
}\kern4pt}} 
\def\gtsim{\hbox{ 
  \vbox to13pt{ 
  \vfill 
  \kern5.5pt 
  \hbox{$>$} 
  \kern-15pt 
  \hbox{$\sim$} 
}\kern4pt}} 

\title{Large fluctuations of disentaglement force and 
implications for polymer dynamics}
\author{Raphael Blumenfeld\inst{1,2}}
\institute{
  \inst{1} Polymers and Colloids - Cavendish Laboratory, 
  Madingley Road, Cambridge CB3 0HE, UK \\
  \inst{2} Physics Department, Warwick University - Coventry CV4 7AL, UK}

\begin{document}

\maketitle

\begin{abstract}
This paper examines the effect of cooling on disentanglement forces in polymers 
and the implications for both single chain pullout and polymer dynamics. 
I derive the explicit dependence of the distribution of these forces on 
temperature, which is found to exhibit a rich behaviour.  
Most significantly, it is shown to be dominated by large fluctuations up to a 
certain temperature $T_0$  that can be determined from molecular parameters. 

The effects of these fluctuations on chain friction are analysed and they are 
argued to undermine the traditional melt-based models that rely on a typical chain 
friction coefficient.  
A direct implication for first principles calculation of viscosity is discussed.
This quantifies the limit of validity of such descriptions, such as Rouse dynamics 
and the Tube model, and pave 
the way to model polymer dynamics around the glass transition temperature. 
\end{abstract}

Polymer chain dynamics in dense melts are described quite well by Rouse dynamics and 
the Tube model \cite{edoi}\cite{degennes} that have become the accepted paradigm. The
latter assumes that a chain moves in an effective tube formed by nearby chains and 
against friction produced by disentanglement from the rest of the network. 
This mean field type of description becomes increasingly inaccurate upon cooling and 
loses validity altogether around the glass transition. 
While there have been attempts to improve the model by considering more than one chain 
(e.g., through constraint release \cite{manychainmodels}), it is worthwhile to explore 
extensions to the one chain picture before rushing to more complicated models. 
A detailed understanding of chain dynamics over a wide range of temperatures is also 
required to interpret recently emerging experiments of single chain pullout \cite{pullout}.
A model for the latter has been proposed recently \cite{bi} in which the chain's mean 
field environment is replaced by a detailed consideration of the local dynamics with
a focus on the distribution of forces needed to disentangle non-permanent entanglement 
points (EPs). 
An intriguing result of reference \cite{bi} is that at low temperatures 
(below the glass transition) this distribution is dominated by large fluctuations. 
This result raises several questions: 
(1) while it is evident that the fluctuations must decay with increasing temperature, 
it is unknown how, and at what rate, this happens, nor is it known to what temperature 
do the fluctuations persist. 
(2) is it possible to quantify the effect of the large fluctuations on rheological 
properties such as viscosity and so determine the range of validity of melt-based models? 
The latter issue is essential as a basis for modelling polymer dynamics around the glass 
transition temperature.  

These issues are the subject of this Letter. The probability density function (PDF) of 
the disentanglement forces and its temperature dependence are derived. 
The PDF is found to exhibit a rich behaviour: 
At high temperatures it is sharply peaked around very small forces, as expected, but  
upon cooling the disentanglement forces increase and a broad local maximum develops. 
Below some temperature a gap appears at the low forces end which gradually increases 
until it becomes complete at very low temperatures. When this happens the PDF diverges 
at a value, $f_0$, which is determined by {\it molecular data}. An 
algebraic tail then appears at the large forces end of the PDF, giving rise to even larger 
fluctuations. 
I discuss the implications of these results for viscosity calculations from molecular 
information and derive the friction force on individual polymer chains. The connection 
between the broadening of the PDF and appearance of large fluctuations in chain friction 
is used to quantify how melt-based models break down in this regime.

We start by deriving the disentanglement threshold force distribution using the model 
introduced in \cite{bi}. 
\begin{figure}
\onefigure{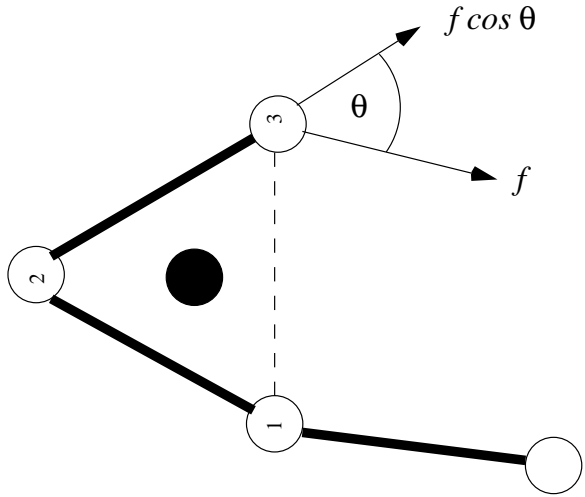}
\caption{A non-permanent entanglement: the black circle is an anchor
perpendicular to the page plane around which the disentangling chain slides.}
\label{f.1}
\end{figure}
In this model a non-permanent EP is regarded as a local wrapping of one
chain around an anchor chain which is fixed by the surrounding network.
Disentanglement occurs when the wrapped chain slides around the anchor  
under a pulling force, as shown in fig.~\ref{f.1}.
The activation energy required to move $m$ segments one position clockwise, $E_b=mE$, is 
determined by the molecular interactions and for anchors consisting of one chain $m=2$ is 
expected (e.g., as shown in fig.~\ref{f.1}). 
The thicker the anchor the higher the value of $E_b$. 
Tugging at bead 3 with an increasing force $f$ in an arbitrary direction exerts 
a force $(f\cos\theta)$ along the 2-3 link and the disentanglement force is the 
value of $f$ when sliding starts. 
For simplicity, it is assumed that once a chain starts slipping it continues to 
do so until disentanglement is complete. Once disentangled, the portion of the chain 
immediately after it starts to uncoil, building stress on the next EP. 
This process continues until the entire chain is unraveled. 
At very low temperature, slipping occurs when the work needed to move the chain by one 
segment $a$ is equal to the activation energy $E_b$, $f\cos\theta = E_b / a$. 
The value of $E_b / k_B$ ($k_B$ being Boltzmann constant) is expected to depend only weakly 
on molecular weight. 
In reference \cite{bi} the force distribution was derived at 
low (essentially zero) temperature. Here we investigate the effect of both 
the distribution of $\theta$ and the thermal fluctuations on the distribution 
of $f$. 

\begin{figure}
\onefigure{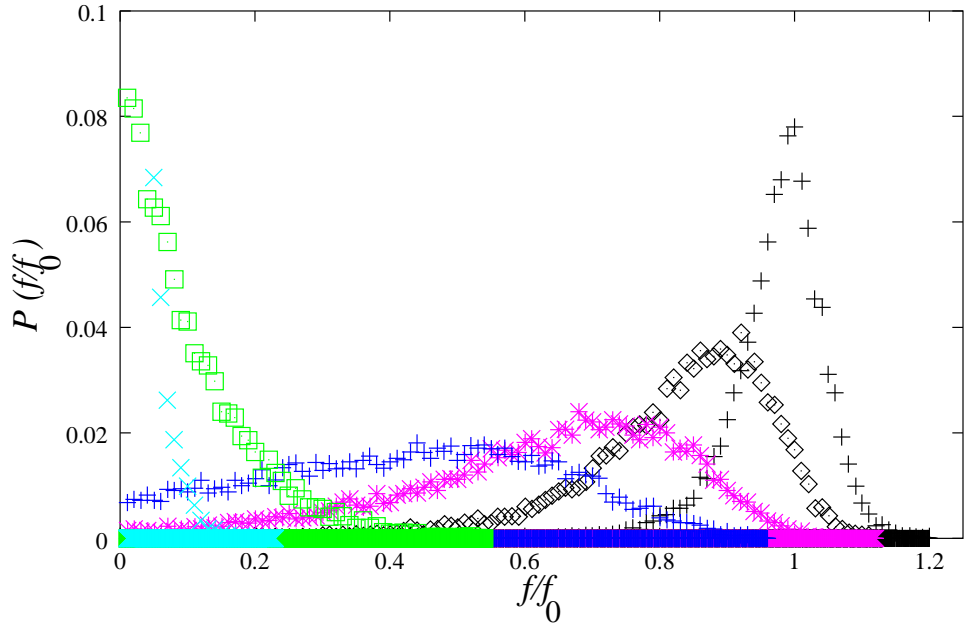}
\caption{The PDF of the disentanglement forces $P_f(f)$ for 
$T = 0.05(+), 0.1(\diamond), 0.15(*), 0.2(+), 0.4(\fbox{\ })$ and $0.9(\times)\ \beta E_b$. 
Note the broadening and the development of a local maximum, which moves to larger 
forces with decreasing temperature.} 
\label{f.2}
\end{figure}

Since equilibrated polymer networks are isotropic down to the scale of a few monomers the angle 
$\theta$ is uniformly distributed but only values of $\theta$ in the range 
$0 < \theta < \theta_{max} \equiv \pi/2 - \eps$ need be considered. This is because:
(i) $\theta<0$ leads to unwrapping rather than sliding, and 
(ii) values of $\theta$ too close to $\pi/2$ are irrelevant because the forces required 
to overcome the threshold $f_0$ in this regime are too large ($\sim 1/\theta$) and would 
cause scission before sliding occurs. It follows that

\begin{equation}
\label{Mi}
\cos\theta_{max} = E_b / (af_{scission}) \equiv \epsilon \ll 1 
\end{equation}
where $f_{scission}$ is the force to rupture the chain and $a$ is Kuhn's length. 
As the chain is pulled, the force on the EP steadily increases and disentanglement occurs 
when the energy input is larger than $(E_b-af\cos\theta)$. 
The probability of disentanglement to occur at all is therefore proportional to 
$e^{-\beta(E_b-af\cos\theta)}$ when $f\cos\theta < f_0 $ and identically unity otherwise.
The probability of the former scenario is

\begin{equation}
\label{Tiii}
P_1 = \int_{\gamma(f)}^{\theta_{max}}
e^{-\beta (E_b - a f\cos\theta)} \left[ 1 -
\int_0^f e^{-\beta (E_b - a f'\cos\theta)} df'\right] d\theta 
\end{equation}
and the probability of the latter one is 

\begin{equation}
\label{Tii}
P_2 = {1\over{f\sqrt{(f/f_0)^2 - 1}}} \left[ 1 - \int_0^f 
e^{-\beta (E_b - a f'\cos(\gamma(f))} df'\right] \delta[\gamma(f)-\theta] \ . 
\end{equation}
In these and the following expressions $\beta = 1/k_B T$, $\gamma(f) = {\rm arccos}(f_0/f)$, 
$\delta(x)$ is Dirac's delta-function and $H(x)$ is the step function 
which is 1 for $x>0$ and 0 otherwise. The PDF that disentanglement occurs exactly 
when the pulling force reaches a value between $f$ and $f+df$ is therefore 
\begin{equation}
\label{Tiv}
P(f,T) df = C (P_1 + P_2) df =  C \left[
{{H(f-f_0)}\over{f \sqrt{\left(f/f_0\right)^2 - 1}}} + 
e^{-\beta E_b} S \right] df 
\end{equation}
where
$$S = - {{H(f-f_0)}\over{\sqrt{\left(f/f_0\right)^2 - 1}}}
{{e^{\beta a f_0} - 1}\over{\beta a f_0}} + 
\int_{\gamma(f)}^{\theta_{max}} e^{\beta a f \cos\theta} \left[ 1 - 
{{e^{\beta a f\cos\theta} - 1}\over{\beta a \cos\theta}} e^{-\beta E_b} 
\right]  d\theta $$
and the normalization constant is
$$C = \left(\theta_{max} + e^{-\beta E_b}
\int_0^{f_0/\epsilon} S df \right)^{-1} \ .$$
When $T\to 0$ $C\to 1/\theta_{max}$ and the PDF converges to the first term, 
reproducing in this limit the result of reference \cite{bi}.  
The second term gives the probability density of thermally assisted 
disentanglement and governs the range $f<f_0$.
Fig.~\ref{f.2} shows $P(f,T)$ generated numerically at  
$T = 0.05, 0.1, 0.15, 0.2, 0.4$ and $0.9$, where 
the temperature is measured in units of $E_b/k_B T$. 

\begin{figure}
\onefigure{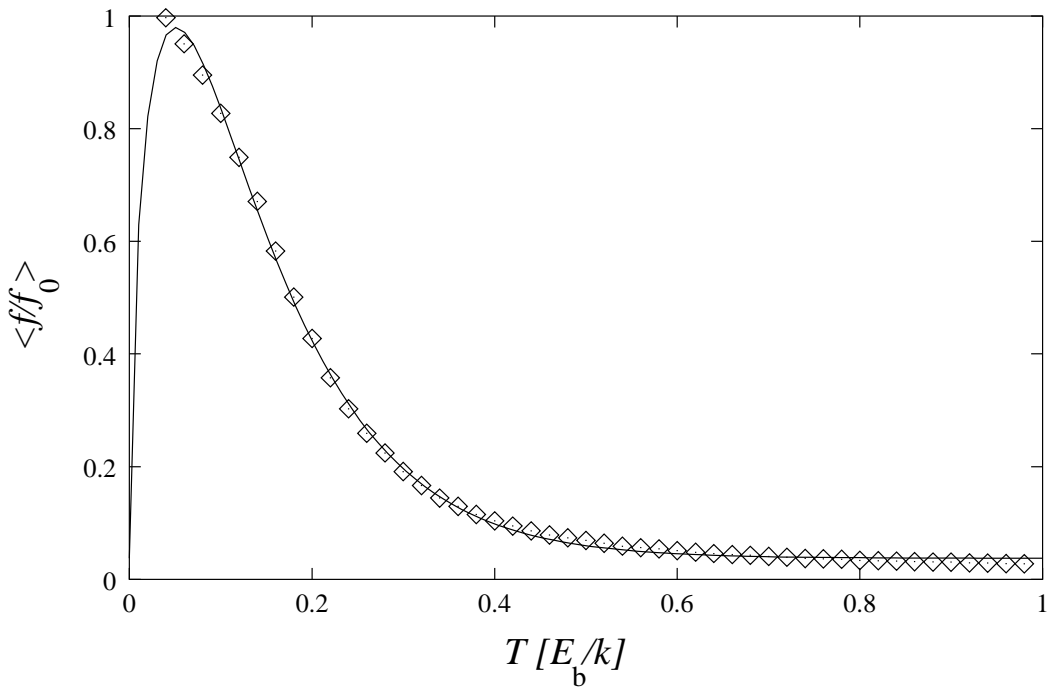}
\caption{The mean disentanglement force as a function of temperature. The 
points are obtained from a numerical calculation and the line is  
a power-exponential fit, eq.~(\ref{Tvi}). Note that this curve is also 
proportional to the viscosity $\xi$, eq.~(\ref{Tv}).  }
\label{f.3}
\end{figure}

At $T=0.9$ the PDF is narrowly peaked around $f\approx 0.05f_0$ and 
as the temperature  decreases it develops a maximum which moves towards 
higher values of $f$. Around $T\approx 0.2$ the PDF is smeared almost over the 
entire range $0 < f < f_0$ with a broad hump around $f\approx 0.5f_0$. 
The maximum sharpens again with further cooling while moving to higher 
values of $f$. Another notable feature is the appearance of a gap at small values of 
$f$ as the temperature drops below $T=0.2$, corresponding to  an exponentially low 
probability of disentanglement by small forces.
Around the same temperature a tail starts growing at values of $f$ larger than $f_0$.
Finally, as $T$ approaches zero, the gap is complete in  
the sense that no force below $f_0$ can cause disentanglement, the PDF is 
sharply peaked at $f_0$ and the tail assumes the algebraic form of 
the first term in eq.~(\ref{Tiv}).
Fig.~\ref{f.3} shows the dependence of the mean force, $F = \langle f/f_0 \rangle$,
on temperature for $N=100$. In principle, $\langle f(T)\rangle$ should be computable 
directly from the PDF but the resulting expressions are cumbersome. 
Empirically, the data can be fitted very well empirically by 
\begin{equation}
\label{Tvi}
\langle f/f_0 \rangle = A T^{x} e^{- y T} + B\ ,
\end{equation}                                                                                     
where $A=9\pm 1$, $x=0.57\pm 0.03$, $y=11.2\pm 0.3$ and $B=0.037\pm 0.002$. 
It is important to note that the values of these parameters  depend on
molecular data through the dimensionless quantity $\epsilon=E_b/(af_{scission})$. 
This has the intriguing implication that changing the chemistry to vary
$E_b$ and $f_{scission}$ can have strong effects on the dynamics and rheology in
the manner described by the results below.
The very low-temperature behavior ($T < 0.1$) is dominated by large fluctuations of the form 
 
\begin{equation}
\label{Tiva}
\sqrt{\langle \delta^2f \rangle} \sim {1\over{\sqrt{\epsilon}}} 
+ O\left(e^{-\beta E_b}\right) \ .
\end{equation}
As the temperature increases the moments of $f$ get smaller, the second term in 
eq.~(\ref{Tiva}) becomes dominant, and the sensitivity to $\epsilon$ is washed out.
Both the long tail at low temperatures and the broad distribution at 
intermediate temperatures suggest that large fluctuations are dominant 
below some temperature $T_0$. Note also that the smaller the value of $\epsilon$ (the 
higher the value of $f_{scission}$) the 
larger the fluctuations and therefore the higher is $T_0$, suggesting a way to  
manipulate $T_0$ via the chemistry.

To estimate the crossover temperature $T_0$ consider the relative width of the PDF, 
$\Gamma \equiv \langle\delta f^2\rangle^{1/2}/\langle f\rangle$.
When $\Gamma\ll 1$ fluctuations are insignificant and the force is well 
characterised by a typical value, $f_{typ}$. When $\Gamma$ is of order 1,
fluctuations dominate and this is not a good approximation.
Figure 4 shows that $\Gamma$ increases steadily with temperature up to $T \approx 0.4$ and 
then saturates to a plateau, $\Gamma_p \approx 0.8$. 
Thus, fluctuations play a significant role at all temperatures up to $T = 1$. However, 
above $T\approx 0.4$ the mean force is quite small and even the fluctuations cannot make it 
significantly large. As shown below, large force fluctuations lead to significant fluctuations 
of the chain friction coefficient, which mark the failure of melt-based models.
Well above $T_0$ thermal noise overwhelms the activation energy, the forces become negligibly 
small and the dynamics should converge to the Tube model. 

\begin{figure}
\onefigure{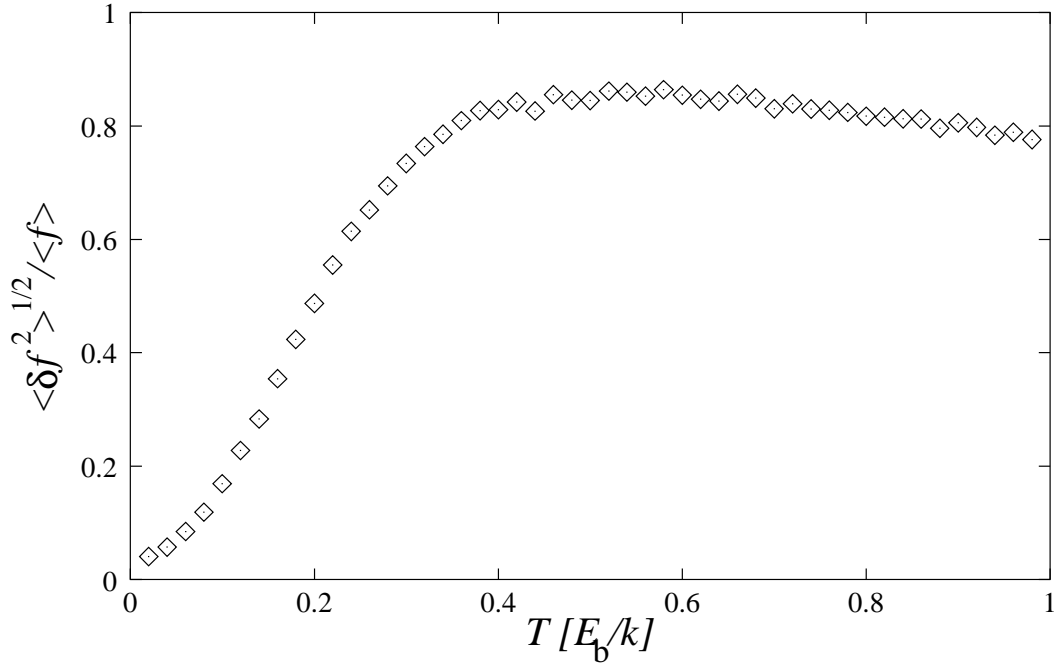}
\caption{The relative width, $\Gamma = \langle\delta f^2\rangle^{1/2}/\langle f \rangle$ 
as a function of temperature.}
\label{f.4}
\end{figure}
To link these results to rheological behaviour, it is prudent to first establish the relation 
between chain friction and the disentanglement force distribution.
Pulling a chain at a rate $v$ consists of alternating uncoiling and disentanglement 
processes, giving rise to an oscillating pulling force. 
As the $(n-1)$th EP dientangles the segment following it uncoils behaving as 
an entropic spring whose effective length is the chain segment between the pulling 
point, $n=0$, and the $n$th EP. Once the force reaches the disentanglement threshold 
of the $n$th EP it disentangles and the force drops. 
Presuming equilibration on the time scale of this process, the drop of the force 
is to the value consistent with elasticity of an entropic spring longer by the segment between 
the $n$th and $(n+1)$th EP. 
Now, the number of EPs disentangled per unit time $\tau$ is $N = (v\tau/l_0)$, where 
$l_0$ is the typical distance along the chain (measured in units of Kuhn's length $a$) 
between two successive EPs \cite{flory}. The force invested is 

\begin{equation}
\label{Tv}
F = \tau^{-1}\sum_{n=1}^N \int_0^{t_n} f_n(t) dt = \langle f \rangle_N
\end{equation}
where $t_n$ is the time between the disentanglements of the $(n-1)$th and
$n$th EPs, $f_n(t)$ is the force during the time interval $t_n$, and 
$\langle ... \rangle_N$ indicates a time average for a given number of EPs $N$. 
To maintain a constant pulling rate, $v$, this force is balanced by friction and 
therefore $\xi v = F$. 
Therefore, fig.~\ref{f.3} gives (up to a numerical factor) $\xi(T)$ for a particular 
value of $N$. 
If the terms on the right hand side of eq.~(\ref{Tv}) have a typical value, $g$, 
then the friction is $g/l_0$. If, however, $f_n$ is broadly distributed then Eq.~(\ref{Tv}) 
makes it immediately apparent that $\xi$ can also suffer large fluctuations and its 
measured value may depend nontrivially on $N$. 
Furthermore, the sensitivity of the distribution of $f_n$ to temperature affects 
directly the temperature dependence of the mean and the fluctuations of $\xi$. 
We therefore expect $\xi$ to be characterised well by a typical value only above $T_0$, where 
the distribution of $f_n$ is narrow. Below $T_0$ its measured value should appear noisy. 
Large fluctuations in $\xi$ undermine melt-based models of polymer dynamics. 
For example, Rouse dynamics are based on 
a Langevin-type equation where the chain friction coefficient is presumed to have a typical 
constant value, and this description breaks down when the fluctuations of $\xi$ become relevant. 
Similarly, the Tube model needs to be reconsidered when the chain friction is widely 
distributed. 
It is therefore suggested here that, in the absence of other mechanisms for the dynamics, 
$T_0\approx 0.4$ is the crossover temperature below which melt-based models for polymer dynamics break down.
Since one expects such models to fail around the glass transition then $T_g\approx T_0$,
suggesting that $T_g$ may be found from {\it molecular data}. 
For example, for polystyrene $T_g\approx 370K$ and therefore $E_b/k_B\approx 950K$ and 
$f_0\approx 50$pN. Thus this conjecture can be tested 
experimentally using the recent techniques of single chains pullout. 

To conclude, I have derived the distribution of forces required to disentangle 
non-permanent entanglement points as a function of 
temperature. I have shown that this distribution is narrowly peaked around small forces 
for temperatures in the melt and it broadens with cooling.
At low temperatures a gap develops which corresponds to the increasing difficulty to 
disentangle at low pulling forces, as one would expect.
The distribution, and consequently the dynamics, have been shown to depend on only 
two molecular quantities: $E_b$, the barrier energy needed to slide one chain 
across another, and $\epsilon=E_b/(a f_{scission})$.
The pulling force fluctuations have been found to dominate the dynamics below 
$T_0\approx 0.4 E_b/k_B$ and the implications of these results to viscosity on the molecular 
level have been discussed.
This analysis pinpoint the range of validity of melt-based models by quantifyings how and at 
what temperature they break down. It also forms a basis to construct a consistent model 
that can bridge between the dynamics in the melt and glass states. 
The present description holds as long as the chain consists of both coils and non-permnanent 
entanglement points and is not valid for too slow an equilibration or too fast a pulling rate.

Since the chain dynamics are fundamental to most aspects of polymer systems these results 
are expected to have a wide range of applications. 
For example, they enable to compute directly the mean force required to pull an ensemble of 
chains along an interface in the interior of a polymer system, which is directly relevant
to yield stresses of glassy polymers. Carrying a similar analysis on interfaces  
between different polymer materials (joined by welding, grafting, or any other method) would 
give the strength of such joints. An application to the relaxation behaviour 
of rubbers is being currently considered by this author. On a fundamental level, it is 
possibloe to re-analyse the Rouse dynamics with a noisy friction coefficient using the results 
obtained here, extending the usefulness of the traditional approach to lower temperatures 
than it is currently valid for. This analysis also paves the way to first principles 
calculations of the viscosity around the glass transition directly from molecular information.

\acknowledgments
It is a pleasure to acknowledge helpful discussions with Prof. Sir S. F. Edwards, 
Dr E. Terentjev and Prof A. M. Donald.

\end{document}